\documentclass[graybox]{svmult}

\usepackage{mathptmx}       
\usepackage{helvet}         
\usepackage{courier}        
\usepackage{type1cm}        
\usepackage{makeidx}         
\usepackage{graphicx}        
\usepackage{multicol}        
\usepackage[bottom]{footmisc}

\usepackage[implicit=false]{hyperref}

\makeindex             

\begin{document}

\title*{Proximity full-text searches of frequently occurring words with a response time guarantee}
\titlerunning{Proximity full-text searches of frequently occurring words}
\author{Alexander B.~Veretennikov}
\institute{Alexander B.~Veretennikov \at Ural Federal University, 620083, 
pr. Lenina 51, Yekaterinburg, Russia, 
Chair of Calculation Mathematics and Computer Science, INSM, 
\email{alexander@veretennikov.ru}}

%
%
\maketitle

\vspace{-3cm}

This is a pre-print of a contribution published in 
Pinelas S., Kim A., Vlasov V. (eds) Mathematical Analysis With Applications. CONCORD-90 2018. Springer Proceedings in Mathematics \& Statistics, vol 318,
published by Springer, Cham. The final authenticated version is available online at: 

\noindent
\href{https://doi.org/10.1007/978-3-030-42176-2\_37}{https://doi.org/10.1007/978-3-030-42176-2\_37}.

\vspace{1cm}

\abstract*{Full-text search engines are important tools for information retrieval. In a proximity full-text search, a document is relevant if it contains query terms near each other, especially if the query terms are frequently occurring words.
For each word in the text, we use additional indexes to store information about nearby words at distances from the given word of less than or equal to $MaxDistance$, which is a parameter. A search algorithm for the case when the query consists of high-frequently used words is discussed. In addition, we present results of experiments with different values of $MaxDistance$ to evaluate the search speed dependence on the value of $MaxDistance$. These results show that the average time of the query execution with our indexes is 94.7--45.9 times (depending on the value of $MaxDistance$) less than that with standard inverted files when queries that contain high-frequently occurring words are evaluated.}

\abstract{Full-text search engines are important tools for information retrieval. In a proximity full-text search, a document is relevant if it contains query terms near each other, especially if the query terms are frequently occurring words.
For each word in the text, we use additional indexes to store information about nearby words at distances from the given word of less than or equal to $MaxDistance$, which is a parameter. A search algorithm for the case when the query consists of high-frequently used words is discussed. In addition, we present results of experiments with different values of $MaxDistance$ to evaluate the search speed dependence on the value of $MaxDistance$. These results show that the average time of the query execution with our indexes is 94.7--45.9 times (depending on the value of $MaxDistance$) less than that with standard inverted files when queries that contain high-frequently occurring words are evaluated.}

\section{Introduction}
\label{sec:Introduction}

   A search query consists of several words. The search result is a list of documents containing these words. In \cite{VeretennikovAB-ProximityFTWithRTG}, we discussed a methodology for high-performance proximity full-text searches and a search algorithm. In this paper, we present an optimization of this algorithm and the results of the experiments in dependence on its primary parameter. 
   
   In modern full-text search approaches, it is important for a document to contain search query words near each other to be relevant in the context of the query, especially if the query contains frequently used words. The impact of the term-proximity is integrated into modern information retrieval models \cite{VeretennikovAB-Yan-Shi-EfficientTPSWithTPI,VeretennikovAB-Buttcher-Clarke-TermProximityScoring,VeretennikovAB-Schenkel-Broschart-EfficientTPS,VeretennikovAB-Rasolofo-Savoy-TermProximityScoring}.
   
\begin{figure}[ht]
    \begin{center}
    \includegraphics[height=5cm]{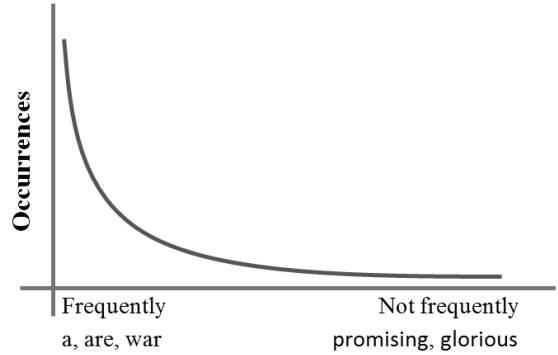}
    \caption{Example of a word frequency distribution.}
    \label{VeretennikovAB_Zipf}
    \end{center}
   \end{figure}

   Words appear in texts at different frequencies. The typical word frequency distribution is described by Zipf's law \cite{VeretennikovAB-Zipf}. An example of words occurrence distribution is shown in Fig. \ref{VeretennikovAB_Zipf}. The horizontal axis represents different words in decreasing order of their occurrence in texts. On the vertical axis, we plot the number of occurrences of each word.
   
   Inverted files or indexes \cite{VeretennikovAB-Zobel-Moffat-InvertedFiles,VeretennikovAB-Tomasic-Etc-IncrementalUpdates} are commonly used for full-text search data structures. With ordinary inverted indexes, for each word in the indexed document, we store in the index the record $(ID, P)$, where $ID$ is the identifier of the document and $P$ is the position of the word in the document (for example, an ordinal number of the word). For proximity full-text searches, we need to store $(ID, P)$ record for all occurrences of any word in the indexed document. These $(ID, P)$ records are called ``postings''. In this case, the query search time is proportional to the number of occurrences of the queried words in the indexed documents. Consequently, it is common for search systems to evaluate queries that contain frequently occurring words (such as ``a'', ``are'', ``war'' and ``who'') much more slowly (see Fig. \ref{VeretennikovAB_Zipf}) than queries that contain less frequently occurring, ordinary words (such as ``promising'' and ``glorious'').
   
   To address this performance problem and to satisfy the demands of the users, we use additional indexes \cite{VeretennikovAB-ProximityFTWithRTG,VeretennikovAB-PhrasesFullText2012,VeretennikovAB-UsingAdditional2013,VeretennikovAB-EfficientFullText2016,VeretennikovAB-CreationAdditional2016,VeretennikovAB-AboutAStructure,VeretennikovAB-EfficientFulltextThreeComponent2017}.
   
   It is important to evaluate any query with a response time guarantee. A full-text search query that we can consider to be a ``simple inquiry'' should produce a response within two seconds \cite{VeretennikovAB-Miller-Response}; otherwise, the continuity of thinking can be interrupted, which will affect the performance of the user.
   
\subsection{Word Type and Lemmatization}
\label{subsec:WordTypeAndLemmatization}

   In \cite{VeretennikovAB-PhrasesFullText2012}, we defined three types of words.
   
   Stop words: Examples include ``and'', ``at'', ``or'', ``not'', ``yes'', ``who'', ``to'', and ``be''. In a stop-words approach, these words are excluded from consideration, but we do not do so. In our approach, we include information about all words in the indexes. We cannot exclude a word from the search because a high-frequently occurring word can have a specific meaning in the context of a specific query \cite{VeretennikovAB-ProximityFTWithRTG,VeretennikovAB-Williams-PhraseQueringCombined}; therefore, excluding some words from consideration can induce search quality degradation or unpredictable effects \cite{VeretennikovAB-Williams-PhraseQueringCombined}. Let us consider the query example ``who are you who''. The Who are an English rock band, and ``Who are You'' is one of their songs. Therefore, the word ``Who'' has a specific meaning in the context of this query.
   
   Frequently used words: These words are frequently encountered but convey meaning. These words always need to be included in the index.
   
   Ordinary words: This category contains all other words.
   
   We employ a morphological analyzer for lemmatization. For each word in the dictionary, the analyzer provides a list of numbers of lemmas (i.e., basic or canonical forms). For a word that does not exist in the dictionary its lemma is the same as the word itself.
    
    We define three types of lemmas: stop lemmas, frequently used lemmas and ordinary lemmas. We sort all lemmas in decreasing order of their occurrence frequency in the texts. This sorted list we call the $FL$-list. The number of a lemma in the $FL$-list is called its $FL$-number. Let the $FL$-number of a lemma $w$ be denoted by $FL(w)$. 
   
   The first $SWCount$ most frequently occurring lemmas are stop lemmas. 
   
   The second $FUCount$ most frequently occurring lemmas are frequently used lemmas. 
   
   All other lemmas are ordinary lemmas. $SWCount$ and $FUCount$ are the parameters. 

   We use $SWCount = 700$ and $FUCount = 2100$ in the experiments presented.

   If an ordinary lemma $q$ occurs in the text so rarely that $FL(q)$ is irrelevant, then we can say that $FL(q) = \sim $. We denote by ``$\sim$'' some large number.
   
   Let us consider the following text, with the identifier $ID1$: ``All was fresh around them, familiar and yet new, tinged with the beauty …''. This is an excerpt from Arthur Conan Doyle's novel ``Beyond the City''.

   After lemmatization: [all] [be] [fresh] [around] [they] [familiar] [and] [yet] [new] [ting, tinge] [with] [the] [beauty].

   With $FL$-numbers: [all: 60] [be: 21] [fresh: 2667] [around: 2177] [they: 134] [familiar: $\sim$] [and: 28] [yet: 632] [new: 376] [ting: $\sim$, tinge: $\sim$] [with: 40] [the: 10] [beauty: $\sim$].

   Stop lemmas: ``all'', ``be'', ``they'', ``and'', ``yet'', ``new'', ``with'', ``the''.

   Frequently used lemmas: ``fresh'', ``around''.

   Ordinary lemmas: ``ting'', ``tinge'', ``beauty'', ``familiar''.

   In this example we can see that some words have several lemmas. The word ``tinged'' has two lemmas, namely, ``ting'' and ``tinge''. Another example is the word ``mine'' that has two lemmas, namely, ``mine'' and ``my'', with $FL$-numbers of 2482 for ``mine'' and 264 for ``my''.

   \subsection{Query Type}
   \label{subsec:QueryType}
   
   Let us define the following query types.
   
\begin{enumerate}   
\item[$QT1$)] All lemmas of the query are stop lemmas.

\item[$QT2$)] All lemmas of the query are frequently used lemmas.

\item[$QT3$)] All lemmas of the query are ordinary lemmas.

\item[$QT4$)] The query contains frequently used and ordinary lemmas; there are no stop lemmas in the query.

\item[$QT5$)] The query contains stop lemmas. The query also contains frequently used and/or ordinary lemmas.
\end{enumerate}
   
   We presented the results of experiments \cite{VeretennikovAB-ProximityFTWithRTG} while showing that the average query execution time with our additional indexes was 94.7 times less than that required when using ordinary inverted files, when $QT1$ queries are evaluated. The experimental query set contained 975 $QT1$ queries, and each was performed three times. The total search time with ordinary inverted indexes was 8 hours 59 minutes. The total search time with our additional indexes was 6 minutes 24 seconds.
   
   Let $MaxDistance$ be a parameter that can take a value of 5 or 7 or even more. In \cite{VeretennikovAB-ProximityFTWithRTG}, we presented the results of experiments with $MaxDistance = 5$.
   
   Before, in \cite{VeretennikovAB-EfficientFullText2016}, we had presented the results of experiments showing that the average number of postings per query with our additional indexes was 51.5 times less than that required when using ordinary inverted files, when queries with $QT2$--$QT5$ types are evaluated (the $QT1$ type is excluded). $MaxDistance = 5$. The experimental query set contained 5955 $QT2$--$QT5$ queries.
   
   In \cite{VeretennikovAB-EfficientFullText2016}, we also presented the results of experiments showing that the average number of postings per query with our additional indexes was 263 times less than that required when using ordinary inverted files, when queries with $QT1$--$QT5$ types are evaluated and when the $QT1$ type search is limited by an exact search (that is, for a $QT1$ query, we find only documents that contain all query words near each other and without other words between, but the query words can be in any order in the indexed document). $MaxDistance = 5$. This limitation we had overcome in \cite{VeretennikovAB-ProximityFTWithRTG,VeretennikovAB-EfficientFulltextThreeComponent2017} by introducing a new type of additional index (three-component key index) for the $QT1$ queries. The experimental query set contained 4500 queries, where 330 are $QT1$ queries and 462 are $QT2$--$QT4$ queries.
   
   In this paper, in a continuation of \cite{VeretennikovAB-ProximityFTWithRTG}, we present the results of experiments for $QT1$ queries when $MaxDistance$ = 5, 7 and 9. With these results, we can evaluate the search speed with three-component key indexes dependent on the value of $MaxDistance$.
   
   We use different additional indexes depending of the type of the query \cite{VeretennikovAB-ProximityFTWithRTG}.

\begin{enumerate}   
\item[$QT1$)] Three-component key $(f, s, t)$ indexes.

\item[$QT2$)] Two-component key $(w, v)$ indexes.

\item[$QT3$)] Ordinary indexes, skipping NSW (near stop words) records \cite{VeretennikovAB-ProximityFTWithRTG}. 

\item[$QT4$)] Ordinary indexes with skipping NSW records \cite{VeretennikovAB-ProximityFTWithRTG} and two-component key indexes.

\item[$QT5$)] Ordinary indexes with NSW records and two-component key indexes. For each frequently used or ordinary lemma in each document, a record ($ID$, $P$, NSW record) is included in the ordinary index. $ID$ is the ordinal number of the document. $P$ is the corresponding word's ordinal number within the document. The NSW record contains information about all stop lemmas occurring near position $P$ (at a distance $\leq$ $MaxDistance$). This information is efficiently encoded \cite{VeretennikovAB-PhrasesFullText2012,VeretennikovAB-UsingAdditional2013,VeretennikovAB-EfficientFullText2016} and allows to take into account any stop lemmas that occurring near $P$. The postings for a lemma in the ordinary index can be stored in two data streams: the first contains $(ID, P)$ records, and the second contains NSW records. In this case, we can skip NSW records when they are not required.
\end{enumerate}   

\section{The Search Algorithm}
\label{sec:TheSearchAlgorithm}

\subsection{The Search Algorithm General Structure}
\label{subsec:TheSearchAlgorithmGeneralStructure}

\begin{figure}[ht]
    \begin{center}
    \includegraphics[height=9cm]{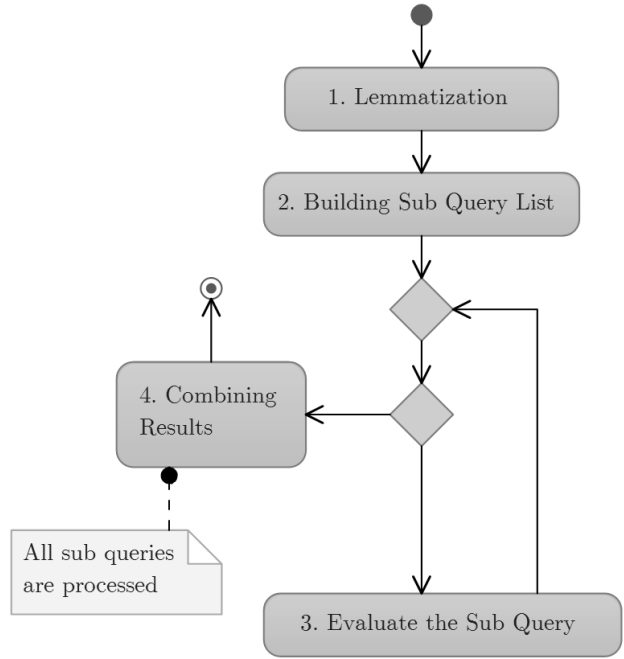}
    \caption{UML diagram of the query evaluation procedure.}
    \label{VeretennikovAB_SearchEn}
    \end{center}
   \end{figure}

   Our search algorithm is described in Fig. \ref{VeretennikovAB_SearchEn}.
   
   Let us consider the following query: ``who are you who''.
   
\begin{table}[ht]
\centering
\caption{The Search Algorithm General Structure.}

\medskip

\begin{tabular}{|l|l|}
\hline
Phase & Result of the phase \\ \hline
1. Lemmatization. & 
The query after lemmatization:
\\ &
[who: 293] [are: 268, be: 21] [you: 47] [who: 293]. \\ \hline
2. Building Sub Query List & 
Q1: [who: 293] [are: 268], [you: 47] [who: 293].
\\ (if required by the query type). &
Q2: [who: 293] [be: 21], [you: 47] [who: 293].
\\ \hline
3. Evaluation of the Sub Queries. & 
Results of $Q1$.
\\ &
Results of $Q2$.
\\ \hline

4. Combining results. &
Combined result set sorted according to relevancy.
\\ \hline
\end{tabular}
\label{tab1}
\end{table}
   
Let us consider the phase 3 in more detail.
We evaluate the sub queries in the loop.
We select a non-processed sub query.
If no such sub query exists, then all sub queries are processed and we go to the next phase.
Otherwise, we evaluate the sub query and go to the start of the loop.

Results of a sub query are the list of records $(ID, P, E, R)$. $ID$ is the identifier of the document. $P$ is the position of the start of the fragment of text within the document that contains the query. $E$ is the position of the end of the fragment of text within the document that contains the query. $R$ is the relevance of the record.

   In \cite{VeretennikovAB-ProximityFTWithRTG}, we defined several query types depending on the types of lemmas they contain and different search algorithms for these query types. In this paper, we consider sub queries that consist only of stop lemmas.
   
\subsection{Evaluation of a Sub Query that Consists only of Stop Lemmas}
\label{subsec:EvaluationOfASubQueryThatConsistsOnlyOfStopLemmas}

   To evaluate a sub query that consists only of stop lemmas, three-component key indexes are used.
   
   The expanded $(f, s, t)$ index or three-component key index \cite{VeretennikovAB-ProximityFTWithRTG} is the list of occurrences of the lemma $f$ for which lemmas $s$ and $t$ both occur in the text at distances less than or equal to $MaxDistance$ from $f$.
   
	For the sub query $Q1$, we can use the (you, are, who) and (you, who, who) indexes. The algorithm for the index selection is described in \cite{VeretennikovAB-ProximityFTWithRTG}.
    
   For each selected index, we need to create the iterator.
   
   The iterator object for the key $(f, s, t)$ is used to read the posting list of the $(f, s, t)$ key from the start to the end.
   
   The iterator object $IT$ has the method $IT.Next$, which reads the next record from the posting list.
   
   The iterator object $IT$ has the property $IT.Value$ that contains the current record $(ID, P)$. Consequently, $IT.Value.ID$ is the $ID$ of the document containing the key, and $IT.Value.P$ is the position of the key in the document.
   
   For two postings $A = (A.ID, A.P)$ and $B = (B.ID, B.P)$, we define that $A < B$ when one of the following conditions are met: $A.ID < B.ID$ or; ($A.ID = B.ID$ and $A.P < B.P$).
   
   The records $(ID, P)$ are stored in the posting list for the given key in increasing order.
   
\begin{figure}[ht]
    \begin{center}
    \includegraphics[height=9cm]{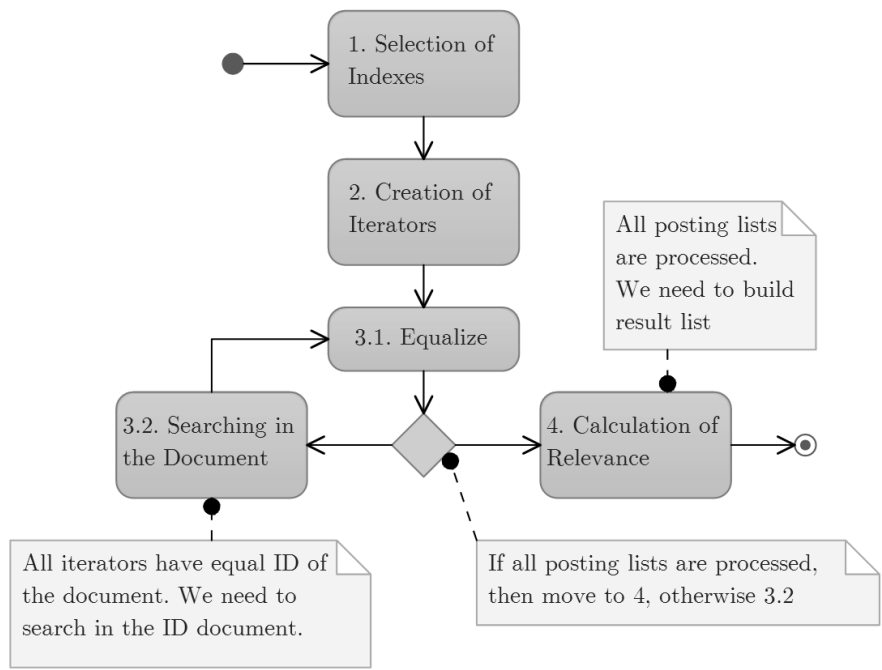}
    \caption{UML diagram of the stop lemma only sub query evaluation procedure.}
    \label{VeretennikovAB_SearchInDocEn}
    \end{center}
   \end{figure}

   The evaluation of the sub query that consists only of stop lemmas \cite{VeretennikovAB-ProximityFTWithRTG} is shown accordingly in Fig. \ref{VeretennikovAB_SearchInDocEn}. Broadly speaking, the evaluation of the sub query is a two level process that is incorporated into the loop (steps 3.1 and 3.2).

\subsection{The Optimized $Equalize$ Procedure}
\label{subsec:TheOptimizedEqualizeProcedure}

\subsubsection{Implementation of $Equalize$ with two Binary Heaps}

   We can implement $Equalize$ with two binary heaps \cite{VeretennikovAB-Williams-Heapsort}. Let $MaxIT$ be the iterator with a maximum value of $Value.ID$. Let $MinIT$ be the iterator with a minimum value of $Value.ID$. If \mbox{$MaxIT.Value.ID = MinIT.Value.ID$}, then all iterators have an equal value of $Value.ID$.
   
   A binary heap is an array of elements $H$. For any elements $A$ and $B$, the comparison operation $A < B$ is defined. This array is indexed from 1.
   
   The binary heap property: for any index $i$, $H[i] \leq H[i \times 2]$ and $H[i] \leq H[i \times 2 + 1]$.
   
\subsubsection{Binary Heap Operations}

   The binary heap provides the following operations.
   
   $Insert(E)$: adds a new element $E$ to the heap with a computational complexity $O(log \: n)$, where $n$ is the count of elements in $H$.
   
   $GetMin$: returns the minimum element with a computational complexity $O(1)$ (returns the first element of the array, i.e., top of the heap). 
   
   $Update(i)$: updates the position of the element with index $i$ with a computational complexity $O(log \: n)$. We will create $H$ as an array of pointers to the iterator objects. Let us consider an example. For any two elements $A$ and $B$ in $H$, we define the operation $A < B$ as $A.Value.ID < B.Value.ID$. Let $IT$ be an element in $H$. When $IT.Next$ is executed, the value of $IT.Value$ is changed, and the position of $IT$ in $H$ must be updated.
   
   We include in any iterator object two additional fields, namely, $MinIndex$ and $MaxIndex$.
   
   We create two heaps, namely, $MinHeap$ and $MaxHeap$.
   
   For $MinHeap$, the operation $A < B$ is defined as $A.Value.ID < B.Value.ID$.
   
   For $MaxHeap$, the operation $A < B$ is defined as $A.Value.ID > B.Value.ID$.
   
   $MinHeap.GetMin$ returns the pointer to an iterator object with the minimum value of $Value.ID$.
   
   $MaxHeap.GetMin$ returns the pointer to an iterator object with the maximum value of $Value.ID$.
   
   In the code for the $Insert$ and $Update$ operations for $MinHeap$ we update the $MinIndex$ field for any iterator object if its position is changed in the heap's array. For any iterator $IT$, the value of $IT.MinIndex$ is always equals to the position of $IT$'s pointer in the $MinHeap$'s array.
   
   In the code for the $Insert$ and $Update$ operations for $MaxHeap$ we update the $MaxIndex$ field for any iterator object if its position is changed in the heap's array. For any iterator $IT$, the value of $IT.MaxIndex$ is always equals to the position of $IT$'s pointer in the $MaxHeap$'s array.
   
\begin{figure}[ht]
    \begin{center}
    \includegraphics[height=75mm]{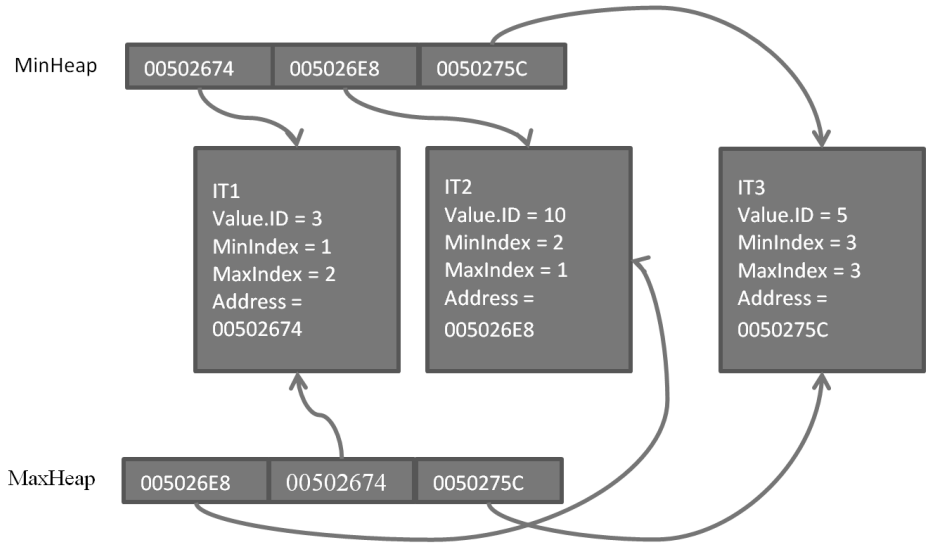}
    \caption{Example of $MinHeap$ and $MaxHeap$ with three iterators.}
    \label{VeretennikovAB_HeapsExample}
    \end{center}
   \end{figure}

   An example of $MinHeap$ and $MaxHeap$ with three iterators is shown in Fig. \ref{VeretennikovAB_HeapsExample}.

   Iterator $IT1$ has $Value.ID = 3$, iterator $IT2$ has $Value.ID = 10$ and iterator $IT3$ has $Value.ID = 5$. 
   
   The $MinHeap$ array has three cells, and the $MaxHeap$ array has three cells.
   
   The $MinHeap$ and $MaxHeap$ arrays contain pointers to the $IT1$, $IT2$ and $IT3$ iterator objects (i.e., the addresses of these objects). To compare two elements of the $MinHeap$ array, we need to obtain two corresponding iterator objects by their addresses and compare their $Value.ID$ fields.
   
   The pointer to the iterator with the minimum value of $Value.ID$, namely, $IT1$, is located in the first cell of the $MinHeap$ array. The pointer to the iterator with the maximum value of $Value.ID$, namely, $IT2$, is located in the first cell of the $MaxHeap$ array.
   
\subsubsection{Details of the Insert Operation}

   For example, in the following code fragment we define the $Insert(IT)$ operation for $MinHeap$. 
   Let $MinHeap.Count$ be the current count of elements in the binary heap $MinHeap$.
   
   Let $MinHeap.Heap$ be the array with length $MinHeap.MaxCount$, indexed from $1$, \mbox{$MinHeap.MaxCount > MinHeap.Count$}.

\begin{enumerate}
\item[1)] $MinHeap.Count = MinHeap.Count + 1$.
\item[2)] $MinHeap.Heap[MinHeap.Count] = IT$.
\item[3)] $IT.MinIndex = MinHeap.Count$.
\item[4)] $i = MinHeap.Count$.
\item[5)] While $i > 1$ and $MinHeap.Heap[i].Value.ID < MinHeap.Heap[i / 2].Value.ID$, perform steps 5.a--5.e.
\begin{enumerate}
\item $T = MinHeap.Heap[i]$, $Q = MinHeap.Heap[i/2]$, 
\item $MinHeap.Heap[i/2] = T$, $MinHeap.Heap[i] = Q$ (swapping $T$ and its parent element).
\item $T.MinIndex = i/2$ (updating $MinIndex$ for $T$).
\item $Q.MinIndex = i$ (updating $MinIndex$ for $Q$).
\item Assignment: $i = i/2$.
\end{enumerate}
\end{enumerate}

   The updating of the $MaxIndex$ field in $MaxHeap$ is performed in a similar way.
   
   We also need to update the $MinIndex$ and $MaxIndex$ fields in the $Update$ operation.
   
\subsubsection{Implementation of $Equalize$}

   We can implement $Equalize$ in the following way.
   
   For any iterator $IT$, we include $IT$ (its pointer) in $MinHeap$ and $MaxHeap$ using $MinHeap.Insert(IT)$ and $MaxHeap.Insert(IT)$. 
   
   Next, in the loop, we perform the following.
   
\begin{enumerate}   
\item[1)] If $MinHeap.GetMin().Value.ID = MaxHeap.GetMin().Value.ID = ID$, then exit from the procedure (for any iterator $IT$ we have $IT.Value.ID = ID$).
\item[2)] Select $IT = MinHeap.GetMin()$.
\item[3)] Execute $IT.Next$.
\item[4)] If no more postings in $IT$, then exit from $Equalize$ and from the search.
\item[5)] Execute $MinHeap.Update(IT.MinIndex)$.
\item[6)] Execute $MaxHeap.Update(IT.MaxIndex)$.
\item[7)] Go to step 1.
\end{enumerate}

\begin{figure}[ht]
    \begin{center}
    \includegraphics[height=11cm]{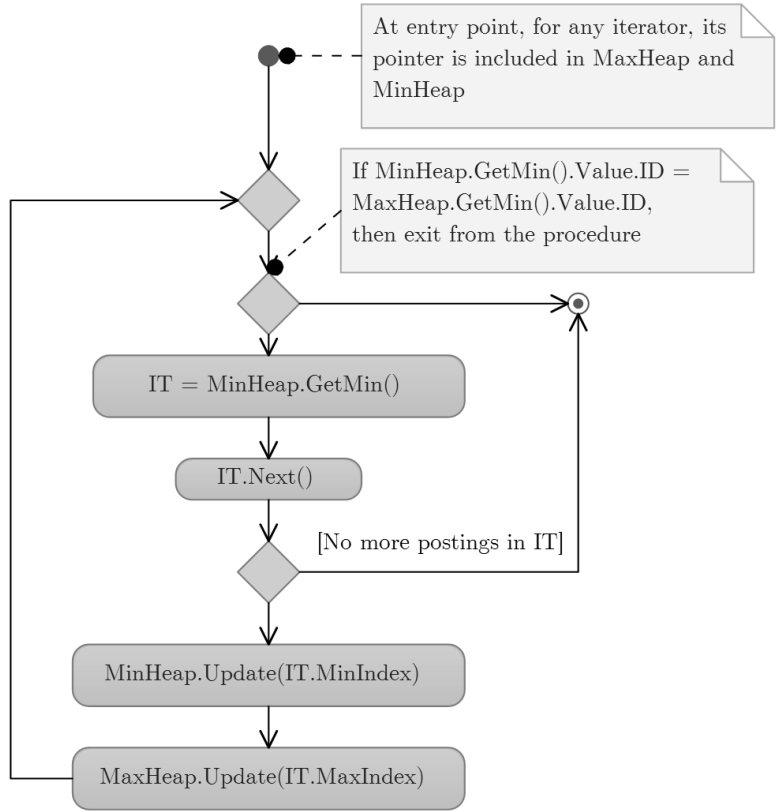}
    \caption{UML diagram of the Equalize procedure.}
    \label{VeretennikovAB_EqualizeHeap}
    \end{center}
   \end{figure}

   The $Equalize$ procedure is shown in Fig. \ref{VeretennikovAB_EqualizeHeap}.

   This implementation of $Equalize$ is more effective and scalable than the basic implementation from \cite{VeretennikovAB-ProximityFTWithRTG} because all operations in the internal loop have a computational complexity $O(log \: n)$, where $n$ is the number of iterators.
   
\section{Search Experiments}
\label{sec:SearchExperiments}

\subsection{Search Experiment Environment}
\label{subsec:SearchExperimentEnvironment}

   In addition to the optimized search algorithm, we discuss the results of search experiments with different values of $MaxDistance$.
   
   All search experiments were conducted using a collection of texts from \cite{VeretennikovAB-ProximityFTWithRTG}. The total size of the text collection is 71.5 GB. The text collection consists of 195 000 documents of plain text, fiction and magazine articles.
   
   $MaxDistance$ = 5, 7 or 9.   
   $SWCount = 700$, $FUCount = 2100$.
   
   The search experiments were conducted using the experimental methodology from \cite{VeretennikovAB-ProximityFTWithRTG}.
   
   We used the following computational resources:
   
   CPU: Intel(R) Core(TM) i7 CPU 920 @ 2.67 GHz.   
   HDD: 7200 RPM. RAM: 24 GB.
   
   OS: Microsoft Windows 2008 R2 Enterprise.
   
   We created the following indexes.
   
   $Idx1$: ordinary inverted file without any improvements such as NSW records \cite{VeretennikovAB-ProximityFTWithRTG}.
   
   $Idx2$: our indexes, including the ordinary inverted index with NSW records and the $(w, v)$ and $(f, s, t)$ indexes, with $MaxDistance = 5$.
   
   $Idx3$: our indexes, including the ordinary inverted index with NSW records and the $(w, v)$ and $(f, s, t)$ indexes, with $MaxDistance = 7$.
   
   $Idx4$: our indexes, including the ordinary inverted index with NSW records and the $(w, v)$ and $(f, s, t)$ indexes, with $MaxDistance = 9$.
   
   Queries performed: 975, all queries consisted only of stop lemmas. The query set was selected as in \cite{VeretennikovAB-ProximityFTWithRTG}. All searches were performed in a single program thread. We searched all queries from the query set with different types of indexes to estimate the performance gain of our indexes.
   
   Query length: from 3 to 5 words. 
   
   Studies by Spink et al. \cite{VeretennikovAB-Spink-AStudy} have shown that queries with lengths greater than 5 are very rare. In \cite{VeretennikovAB-Spink-AStudy}, query logs of a search system were analyzed, and it was established that queries with a length of 6 represent approximately 1\% of all queries and fewer than 4\% of all queries had more than 6 terms.
   
\subsection{Search Experiments}
\label{subsec:SearchExperiments}

   Average query times:
   
   $Idx1$: 31.27 sec., $Idx2$: 0.33 sec., $Idx3$: 0.45 sec., $Idx4$: 0.68 sec.
   
   Average data read sizes per query:
   
   $Idx1$: 745 MB, $Idx2$: 8.45 MB, $Idx3$: 13.32 MB, $Idx4$: 23,89 MB.
   
   Average number of postings per query:
   
   $Idx1$: 193 million, $Idx2$: 765 thousands, $Idx3$: 1.251 million, $Idx4$: 1.841 million.
   
\begin{figure}[ht]
   \sidecaption[t]
    \includegraphics[height=32mm]{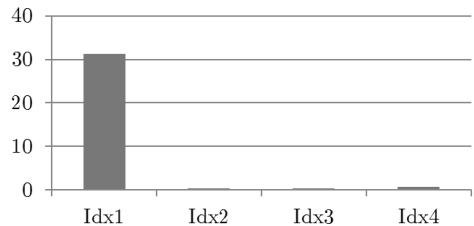}
    \caption{Average query execution times for $Idx1$, $Idx2$, $Idx3$, and $Idx4$ (seconds).}
    \label{VeretennikovAB_AvgSearchTimeIdx14}
   \end{figure}

   We improved the query processing time by a factor of 94.7 with $Idx2$, by a factor of 69.4 with $Idx3$, and by a factor of 45.9 with $Idx4$ (see Fig. \ref{VeretennikovAB_AvgSearchTimeIdx14}).
   
   The left-hand bar shows the average query execution time with the standard inverted indexes. The subsequent bars show the average query execution time with our indexes with $MaxDistance$ = 5, 7 and 9. Our bars are much smaller than the left-hand bar because our searches are very quick.
   
\begin{figure}[ht]
    \sidecaption[t]
    \includegraphics[height=32mm]{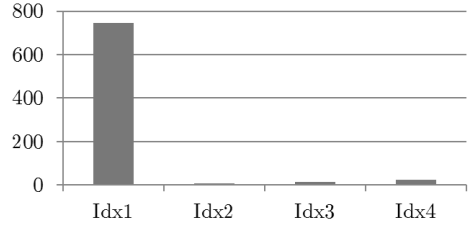}
    \caption{Average data read sizes per query for $Idx1$, $Idx2$, $Idx3$, and $Idx4$ (MB).}
    \label{VeretennikovAB_AvgDataReadSizeIdx14}
   \end{figure}

   We improved the data read size by a factor of 88 with $Idx2$, by a factor of 55.9 with $Idx3$, and by a factor of 31.1 with $Idx4$ (see Fig. \ref{VeretennikovAB_AvgDataReadSizeIdx14}).
   
\begin{figure}[ht]
    \sidecaption[t]
    \includegraphics[height=32mm]{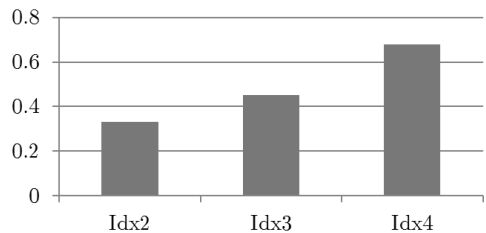}
    \caption{Average query execution times for $Idx2$, $Idx3$, and $Idx4$ (seconds).}
    \label{VeretennikovAB_AvgSearchTimeIdx24}    
   \end{figure}

   We present the differences in the average query execution time for $Idx2$, $Idx3$ and $Idx4$ in Fig. 8 to analyze how the average query execution time depends on the value of $MaxDistance$ (see Fig. \ref{VeretennikovAB_AvgSearchTimeIdx24}).
      
   The left-hand bar shows the average query execution time with $MaxDistance=5$. The subsequent bars show the average query execution time with $MaxDistance = 7$ and 9.

   The search with $Idx3$ was slower than that with $Idx2$ by a factor of 1.36, and the search with $Idx4$ was slower than that with $Idx2$ by a factor of 2.06.

\begin{figure}[ht]
    \sidecaption[t]
    \includegraphics[height=32mm]{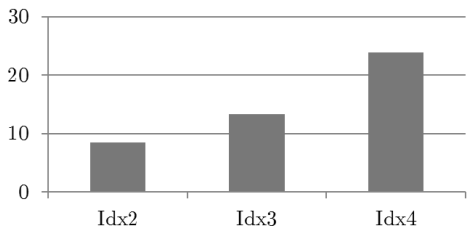}
    \caption{Average data read size per query for $Idx2$, $Idx3$, and $Idx4$ (MB).}
    \label{VeretennikovAB_AvgDataReadSizeIdx24}
   \end{figure}

   We present the differences in the average data read size per query for $Idx2$, $Idx3$ and $Idx4$ in Fig. 9 to analyze how the average data read size depends on the value of $MaxDistance$ (see Fig. \ref{VeretennikovAB_AvgDataReadSizeIdx24}).
  
   The left-hand bar shows the average data read size per query with $MaxDistance=5$. The subsequent bars show the average data read size per query with $MaxDistance = 7$ and 9.
   
   We needed to read from the disk when searching with $Idx3$ more than with $Idx2$ by a factor of 1.57. We needed to read from the disk when searching with $Idx4$ more than with $Idx2$ by a factor of 2.82.
   
\section{Conclusion and Future Work}
\label{sec:ConclusionandFutureWork}

   A query that contains high-frequently occurring words induces performance problems. These problems are usually solved by the following approaches.
\begin{enumerate}
\item[1)] Vertical and/or horizontal increases in the computing resources and the parallelization of the query execution.
\item[2)] Stop words approach.
\item[3)] Early termination approaches \cite{VeretennikovAB-Anh-VectorSpace,VeretennikovAB-Garcia-AccessOrdered}.
\item[4)] Next-word and partial phrase auxiliary indexes for an exact phrase search \cite{VeretennikovAB-Williams-PhraseQueringCombined,VeretennikovAB-Bahle-Auxiliary}.
\end{enumerate}

	The stop words approach leads to search quality degradation \cite{VeretennikovAB-ProximityFTWithRTG} because in some queries a high frequently occurring word can have a specific meaning \cite{VeretennikovAB-ProximityFTWithRTG,VeretennikovAB-Williams-PhraseQueringCombined}, and skipping such a word could lead to the omission of important search results.
    
	Early termination approaches have trouble integrating proximity into the relevance \cite{VeretennikovAB-ProximityFTWithRTG}.
    
	Next-word and partial phrase indexes work only for exact phrase searches.
    
	Our approach allows us to solve performance problems without increasing computing resources, and we can process any word in the query and perform arbitrary queries; these are our advantages.
    
   In this paper, we have introduced an optimized method for full-text searches in comparison with \cite{VeretennikovAB-ProximityFTWithRTG}.
   
   In this paper, we investigated searches with queries that contain only stop lemmas. Other query types are studied in \cite{VeretennikovAB-EfficientFullText2016}.
   
   We studied the dependence of the query execution time on the value of the parameter $MaxDistance$. The results of the search experiments with $MaxDistance = 5, 7$, and 9 are presented. We also proved that a three-component key index can be created with a relatively large value of $MaxDistance = 9$ to allow the effective execution of queries with a length of up to 9 (larger queries need to be divided into parts).
   
   We have presented the results of experiments showing that, when queries contain only stop lemmas, the average time of the query execution with our indexes is 94.7--45.9 times less (with a value of $MaxDistance$ from 5 to 9) than that required when using ordinary inverted indexes. 
   
   When we discuss our indexes, we have shown that with an increase in the value of $MaxDistance$ from 5 to 7, the average query execution time increases 1.36 times. We have shown that with an increase in $MaxDistance$ from 5 to 9, the average query execution time increases 2.06 times. The increase in $MaxDistance$ has a significant impact when we are searching queries that contain only stop lemmas with three component key indexes, but it is still much faster than a search with the standard inverted indexes (improved by a factor of 45.9 for $MaxDistance = 9$). 
   
   In the future, it will be interesting to investigate other types of queries in more detail and to optimize index creation algorithms for larger values of $MaxDistance$.
   
\begin{acknowledgement}
The work was supported by Act 211 Government of the Russian Federation, contract no. 02.A03.21.0006.
\end{acknowledgement}

%
%
%

\end{document}